\documentclass{paper}
\usepackage{epsfig, graphicx}
\usepackage[english]{babel}

\begin{document}
\title{Kinetics of the Phospholipid Multilayer Formation at the Surface of the Silica Sol}
\author{\small Aleksey M. Tikhonov$^{\diamond}$\/\thanks{tikhonov@kapitza.ras.ru}, Viktor E. Asadchikov$^{\dagger}$, Yuri O. Volkov$^{\dagger}$, Boris S. Roshchin$^{\dagger}$,
\\ \small Ivan S. Monakhov$^{\ddagger}$ , and Igor S. Smirnov$^{\ddagger}$}
\maketitle
\leftline{\it $^{\diamond}$ Kapitza Institute for Physical Problems, Russian Academy of Sciences,}
\leftline{\it ul. Kosygina 2, Moscow, 119334, Russia}
\leftline{\it $^{\dagger}$ Shubnikov Institute of Crystallography, Russian Academy of Sciences,}
\leftline{\it Leninskii pr. 59, Moscow, 119333 Russia}
\leftline{\it $^{\ddagger}$ National Research University Higher School of Economics,}
\leftline{\it ul. Myasnitskaya 20, Moscow, 101000 Russia}

\rightline{\today}

\abstract{The ordering of a multilayer consisting of DSPC bilayers on a silica sol substrate is studied within the model-independent approach to the reconstruction of profiles of the electron density from X-ray reflectometry data. It is found that the electroporation of bilayers in the field of anion silica nanoparticles significantly accelerates the process of their saturation with Na$^+$ and H$_2$O, which explains both a relatively small time of formation of the structure of the multilayer
of $(1-7)\times 10^{5}$\,s and 13\% excess of the electron density in it.}

\vspace{0.25in}

\large

A bilayer of phospholipid molecules is considered
as the simplest model of a cell membrane [1-5]. We
previously observed the crystallization of a multilayer
of phospholipid bilayers whose thickness is given by
the Debye screening length $\Lambda_D$ in the bulk of a hydrosol
substrate on the surface of the aqueous solution of
amorphous silicon dioxide nanoparticles \cite{6, 7}
(Fig. 1). In this work, the ordering of the multilayer is
studied within the model-independent approach to
the reconstruction of profiles of the electron density
from X-ray reflectometry data without any a priori
assumptions on the structure of the multilayer [8-12].
According to our data, the characteristic time of formation
of the structure of the surface is $(1-7)\times 10^{5}$\,s;
after that, the lipid film can be considered as a two-dimensional
organic crystal with a quite high degree of
perfection

We study multilayers of 1,2-distearoyl-sn-glycero-
3-phosphoholine (DSPC) or C$_{44}$H$_{88}$NO$_8$P \cite{2}. The
hydrophobic part of the DSPC molecule consists of
two hydrocarbon chains of 18 carbon atoms and has a
length of $\approx 2$\,nm, whereas the hydrophilic part consists
of glycerol and phosphocholine and has a length of
$\approx 1.5$\,nm.

Concentrated monodisperse hydrosols of SiO$_2$
nanoparticles stabilized by sodium hydroxide — Ludox
SM-30 (with the diameter of nanoparticles $\sim 7$\,nm and
the weight fractions of SiO$_2$ and NaOH of 30 and
0.5 wt \%, respectively, and ) and Ludox HS-40
(with the diameter of nanoparticles $\sim 12$\,nm and the
weight fractions of SiO$_2$ and NaOH of 40 and 0.4 wt \%,
respectively, and ) — were used as substrates
[13-16]. The DSPC synthetic phospholipid and silica
hydrosols were purchased from Avanti Polar Lipids
Inc. and Grace Davison Co., respectively.

\begin{figure}
\hspace{0.5in}
\epsfig{file=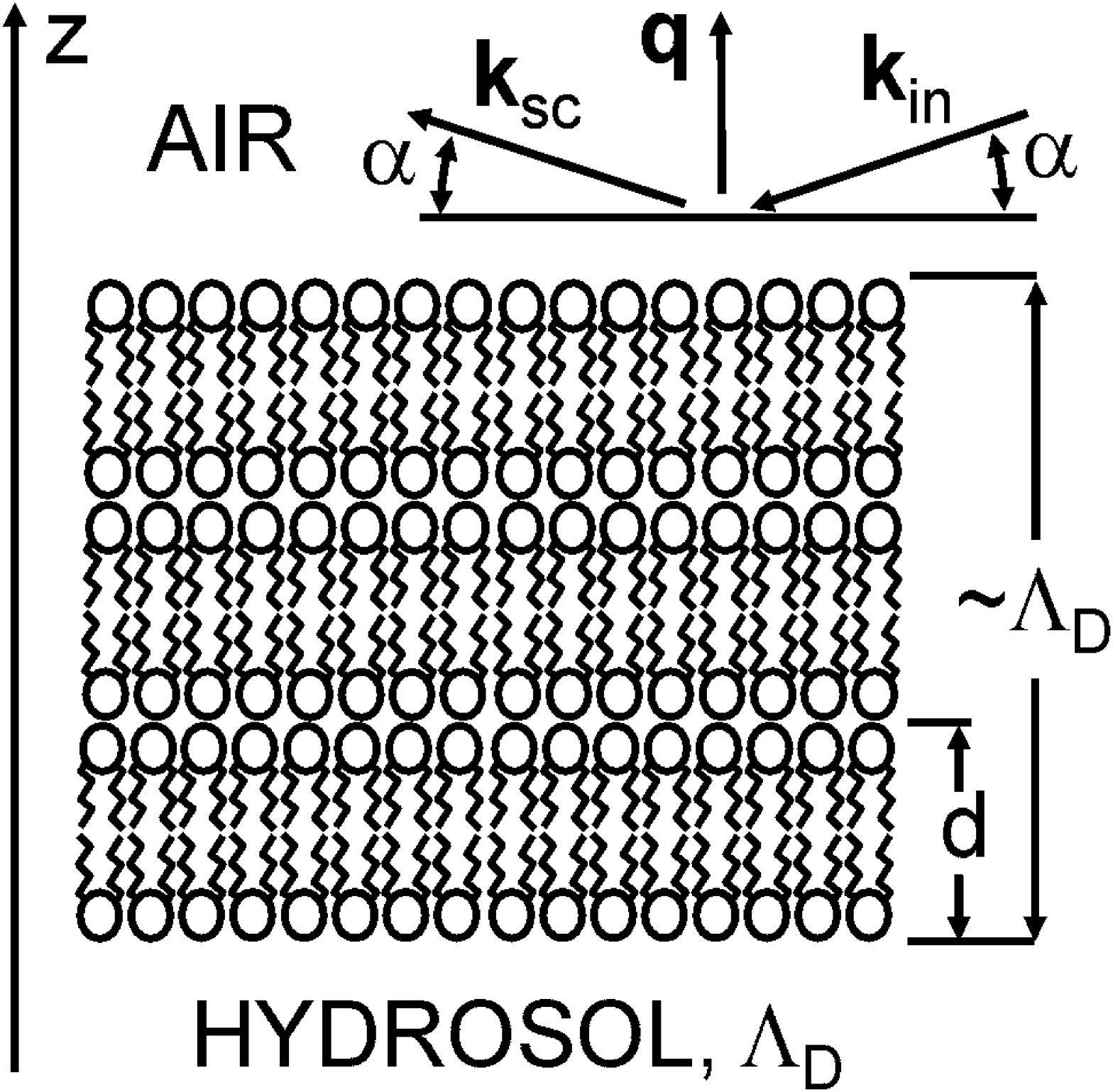, width=0.75\textwidth}

Figure 1. Multilayer of phospholipid bilayers.
\end{figure}

The used silica solutions had the parameter \newline
$\Lambda_D = \sqrt{\epsilon_0\epsilon k_B T/(N_Ae^2c^{-})}$ $\approx 400$\,\AA, where
$\epsilon_0 \approx 8.85\cdot 10^{-12}$\,F/m is the permittivity of free space, $\epsilon \approx 80$ is
the dielectric constant of water, $k_B \approx 1.38\cdot10^{-23}$\,J/K
is the Boltzmann constant, $T \approx 298$\,K is the temperature, $N_A \approx 6.02 \cdot 10^{23}$\,mol$^{-1}$ is the Avogadro number, $e \approx 1.6 \cdot 10^{-19}$\,C is the elementary charge, and $c^{-}\approx 10^{-4}$\,mol/L is the concentration of free OH$^-$ ions
in the sol at \cite{17, 18}. Because of a small difference in pH, the parameter $\Lambda_D$ for
the solution of 12-nm silica particles is larger than that for the solution
of 7-nm particles by $\sim 40\%$.

According to the size distribution of nanoparticles
found from the intensity of small-angle scattering, the
characteristic diameter of particles in the solution is $\sim 30\%$
larger than the value declared by the manufacturer \cite{19, 20}.

\begin{figure}
\hspace{0.5in}
\epsfig{file=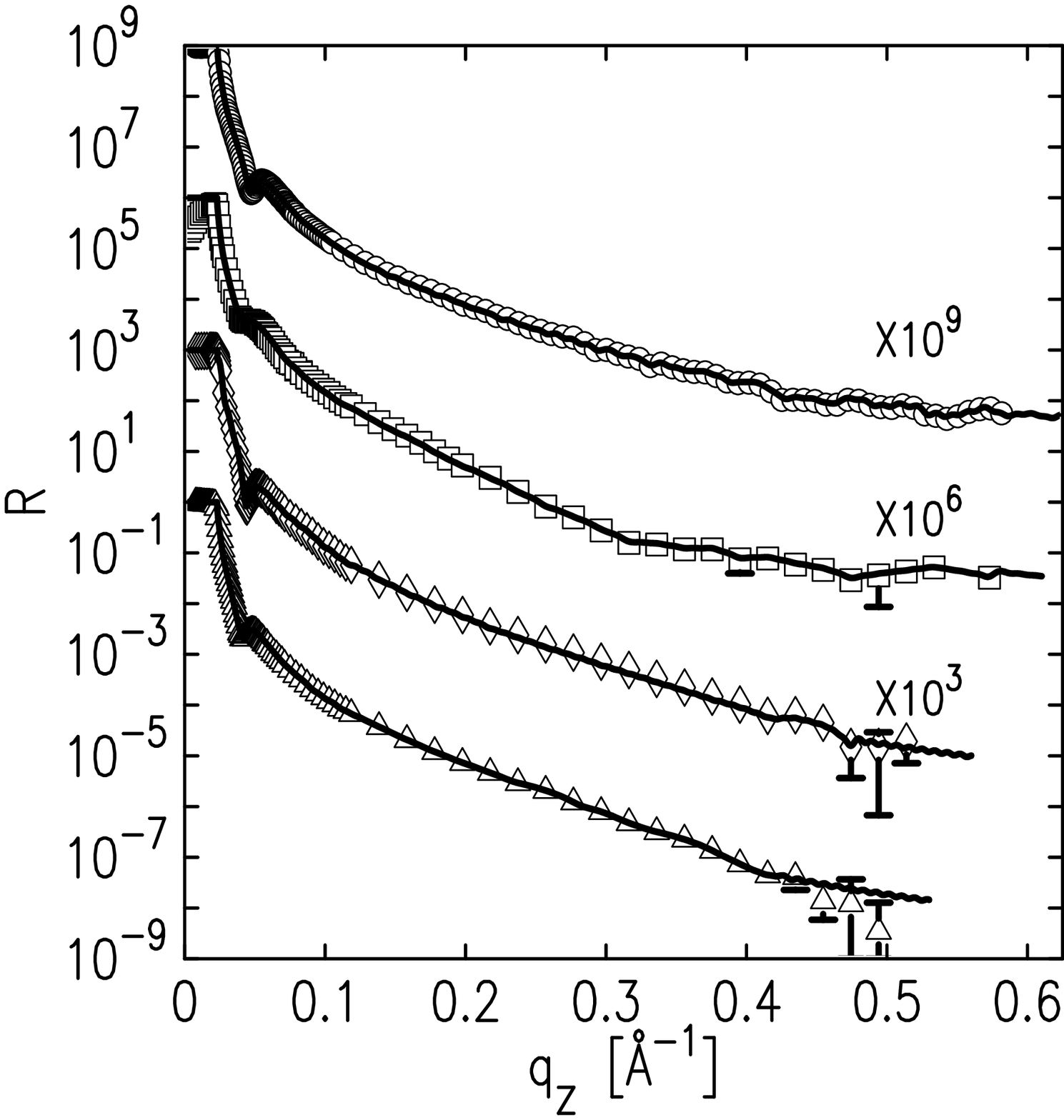, width=0.7\textwidth}

Figure 2. Dependence $R(q_z)$ for the silica sol surface: circles
correspond to the solution of 7-nm particles immediately
after the preparation of the sample, squares are data for this
solution after $\sim 170$\,h, diamonds correspond to the solution
of 12-nm particles immediately after the preparation
of the sample, and triangles are data for this solution after
$\sim 130$\,h. The lines are fits for the reconstruction of the
electron density profiles.
\end{figure}

The multilayer samples were prepared and studied
in a air-tight cell with X-ray-transparent windows
according to the method described in \cite{6}. One or two
drops of the solution of phospholipid in chloroform
($\sim 5 \cdot 10^{-2}$\,mol/L) with a total volume of 10\,$\mu$L were
deposited by a syringe on the surface of the liquid
freshly prepared hydrosol substrate placed in a
polytetrafluoroethylene dish with a diameter of
100\,mm. In this manner, a multilayer of 10-20 lipid
monolayers can be formed on the surface. At such
deposition of the surfactant, its excess is accumulated
in three-dimensional aggregates in equilibrium with
the phospholipid film.

\begin{figure}
\hspace{0.5in}
\epsfig{file=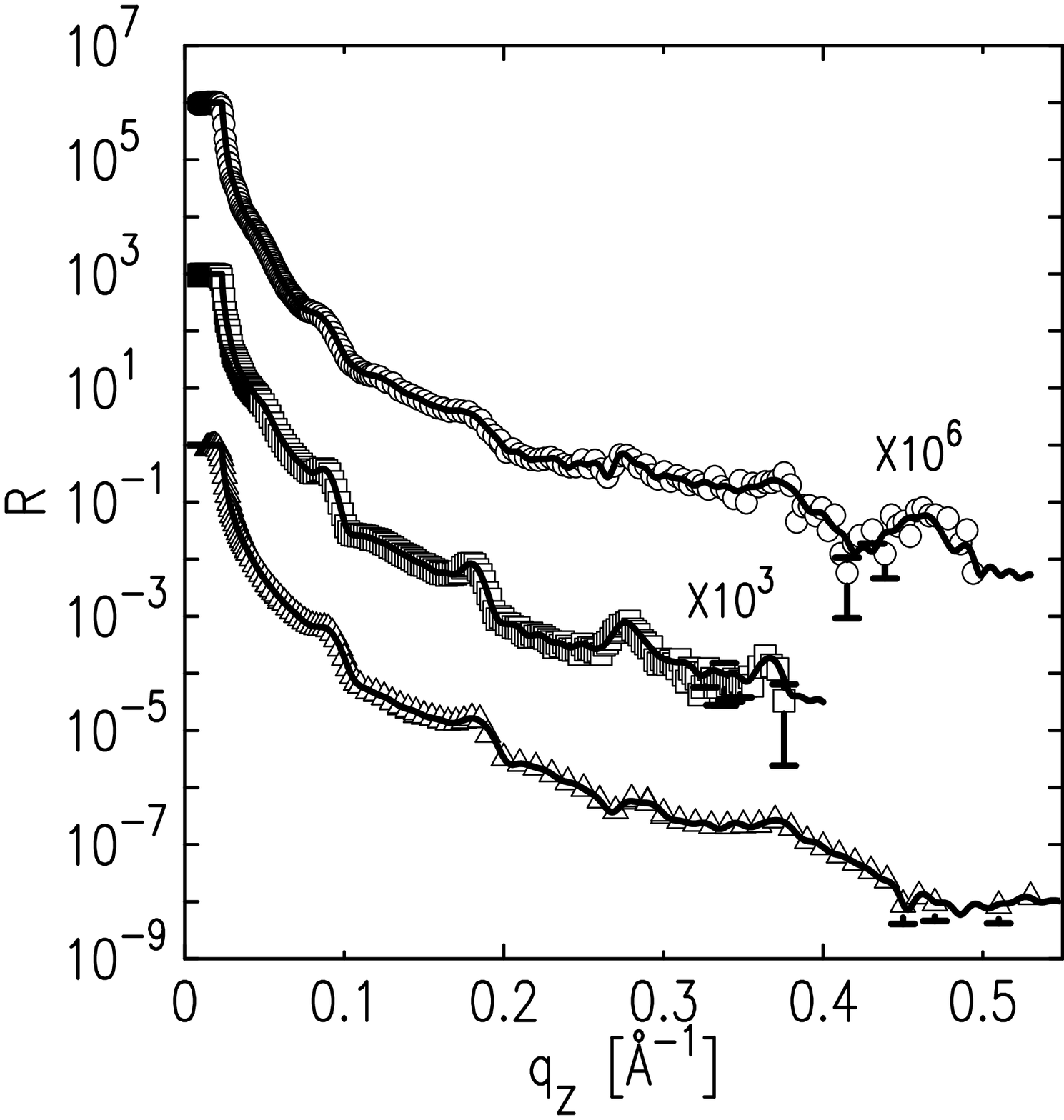, width=0.7\textwidth}

Figure 3. Reflectivity $R(q_z)$ from the DSPC film on the
surface of the solution of 7-nm particles measured (circles)
immediately after deposition, (squares) after $\sim 24$\,h, and
(triangles) after $\sim 100$\,h. The lines are fits for the reconstruction
of the electron density profiles.
\end{figure}

The transverse structure of the lipid layer was studied by
the X-ray reflectometry method on a multipurpose
laboratory diffractometer with a mobile emitter–
detector system \cite{21}.
An X-ray tube with a copper
anode was used as the emitter.
The line $K\alpha_{1}$ (photon
energy $E = 8048$\,eV and wavelength $\lambda = 1.5405 \pm 0.0001$\,\AA{})
 was separated from the tube
radiation spectrum by means of a single Si (111) crystal
monochromator. The vertical and horizontal dimensions
of the beam were $\sim 0.1$ and $\sim 8$\,mm, respectively.
The three-slit collimation system forms a probe X-ray
beam with the angular width in the plane of incidence
$\sim 10^{-4}$\,rad. The angular resolution of the point detector
was $\sim 1.7 \cdot 10^{-3}$\,rad and is determined by the input slit
with a gap of 1\,mm at a distance of$ \sim 570$\,mm from the
center of the sample. To reduce the absorption and
scattering of radiation in air, we used vacuum paths
with X-ray-transparent windows.

At specular reflection, the scattering vector
{\bf q = k$_{\rm in}$ {\rm -} k$_{\rm sc}$}, where {\bf k}$_{\rm in}$ and {\bf k}$_{\rm sc}$ are the wave vectors of
the incident and scattered rays in the direction to the
observation point, respectively, has only one nonzero
component $q_z=(4\pi/\lambda)\sin\alpha$, where $\alpha$ is the glancing
angle in the plane normal to the surface (see Fig. 1).
The software of the diffractometer allows specifying a
variable angular step, width of the slit of the detector,
and time of exposure, which makes it possible to optimize
the measurement of the reflection coefficient $R$,
which decreases rapidly with an increase in $\alpha$.

\begin{figure}
\hspace{0.5in}
\epsfig{file=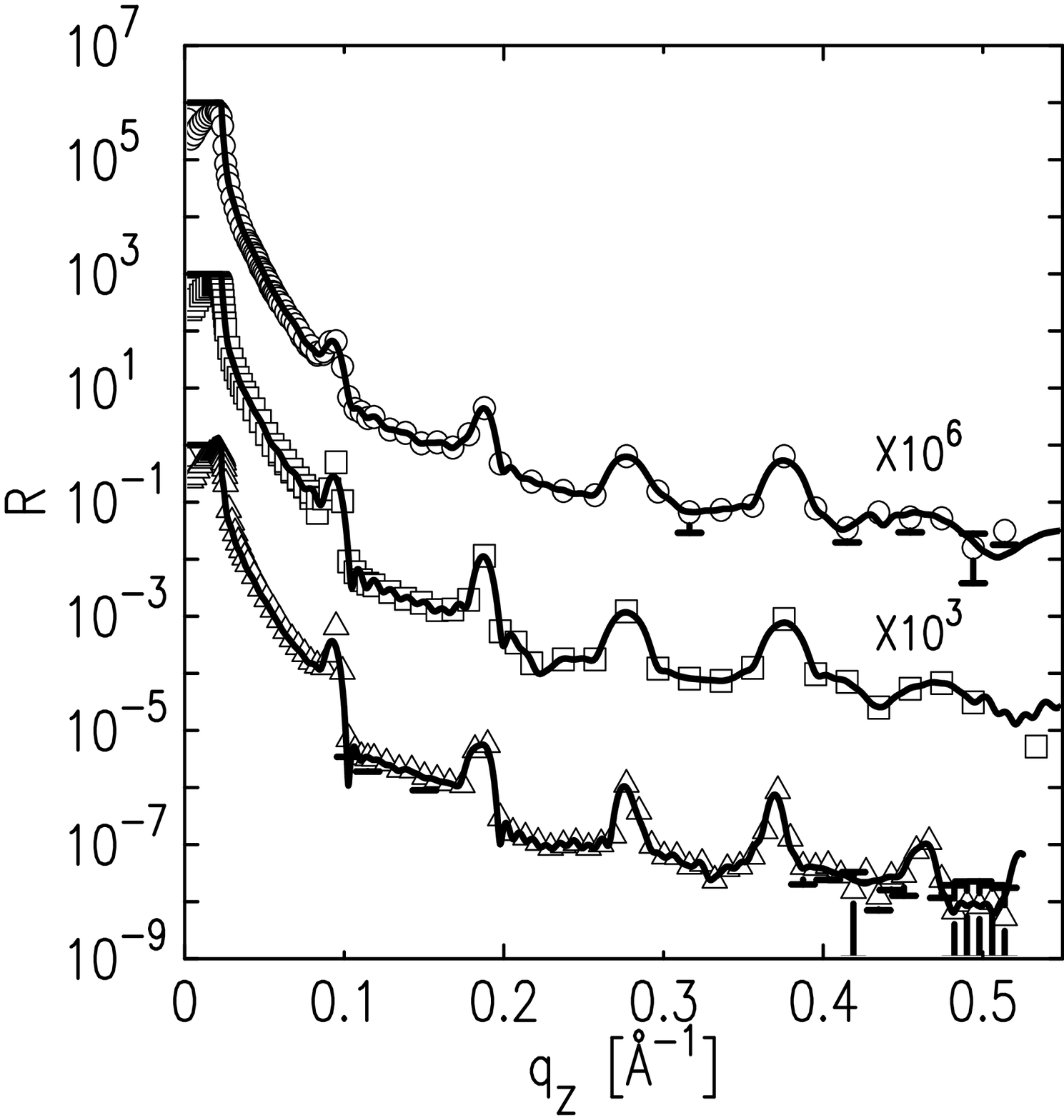, width=0.7\textwidth}

Figure 4. Reflectivity $R(q_z)$ from the DSPC film on the
surface of silica sol of 12-nm particles measured after (circles)
$\sim 24$\,h, (squares) $\sim 70$\,h , and (triangles)$\sim 200$\,h. The lines
are fits for the reconstruction of the electron density profiles.

\end{figure}

Experimental data were processed with a correction
to the shape of the probe X-ray beam because only
a part of the beam reaches the sample surface in the
region of small angles. The background begins to significantly
affect the angular dependence of the reflection
coefficient at large angles. This effect was taken
into account by the subtraction of the previously measured
average value in this angular range, which was $\sim 0.1$~pulses/s.
Thus, the data for the reflection coefficient
obtained on the diffractometer are comparable
in spatial resolution $2\pi/q_z^{max} \approx 10$\,\AA{} (where
$q_z^{max} \approx 0.5$\,\AA$^{-1}$ is the maximum value in the experiment)
with the data previously obtained with synchrotron
radiation \cite{6,7}.

Figure 2 shows the experimental dependences of
the reflectivity $R(q_z)$ from the surface of
clean boundaries of silica sols of (circles, squares) 7-nm
and (diamonds, triangles) 12-nm particles. All curves
exhibit a characteristic feature near $q_z\approx 0.05$~\AA{}$^{-1}$,
which is due to the separation of the components of
the substrate at the (air–silica sol) interface \cite{22}.

Figure 3 shows data for the reflectivity $R(q_z)$
from the DSPC phospholipid film deposited on
the substrate of the silica sol of 7-nm particles. Circles
correspond to reflection from the sample during the
first hour. The characteristic feature near $q_z\approx 0.05$~\AA$^{-1}$
has an inverted shape: a weakly pronounced maximum
is observed instead of the minimum intensity, which
indicates a significant rearrangement of the separation
region. The squares show the reflection coefficient
from the same sample $\sim 24$\,h after the deposition of the
lipid. This curve exhibits a regular set of reflection
peaks with the oscillation period
$\Delta q_z=0.094\pm 0.007$\,\AA{}$^{-1}$, which corresponds to an ordered
structure with an estimated period of
$2\pi/\Delta q_z=66.8\pm 4.5$\,\AA{}. The triangles show the reflection coefficient
from the DSPC lipid multilayer aged for
$\sim 100$\,h. The shape of the reflection peaks and their
period $\Delta q_z=0.096\pm 0.012$\,\AA{}$^{-1}$ almost coincide
with the shape and period of the peaks on the preceding
curve.

Figure 4 shows the dependences $R(q_z)$ for the thick
DSPC phospholipid multilayer which was formed on
the substrate of the silica sol of 12-nm particles. The
measurements were performed at (circles) $\sim 24$\,h,
(squares) $\sim 70$\,h, and (triangles) $\sim 200$\,h after the
preparation of the sample. As in the preceding case,
regular reflection peaks narrowing with the time with
a step of $\Delta q_z=0.096\pm 0.009$\,\AA{}$^{-1}$ are seen on all
curves.

At the measurement of $R(q_z)$, the contribution of
lateral inhomogeneities of the structure is statistically
averaged over the illumination spot on the sample
whose characteristic area was $S \sim 100$\,mm$^2$. For this
reason, the structure of the surface layer of the samples
can be considered in the approximation of an ideal
inhomogeneous-layered structure. To analyze the
reflectometry data, we used the model-independent
approach proposed by Kozhevnikov \cite{8,11}. A significant
advantage of this approach is that it does not
require any a priori assumptions on the shape of the
structure under study, in contrast to the widely used
model approach [23-27].

\begin{figure}
\hspace{0.5in}
\epsfig{file=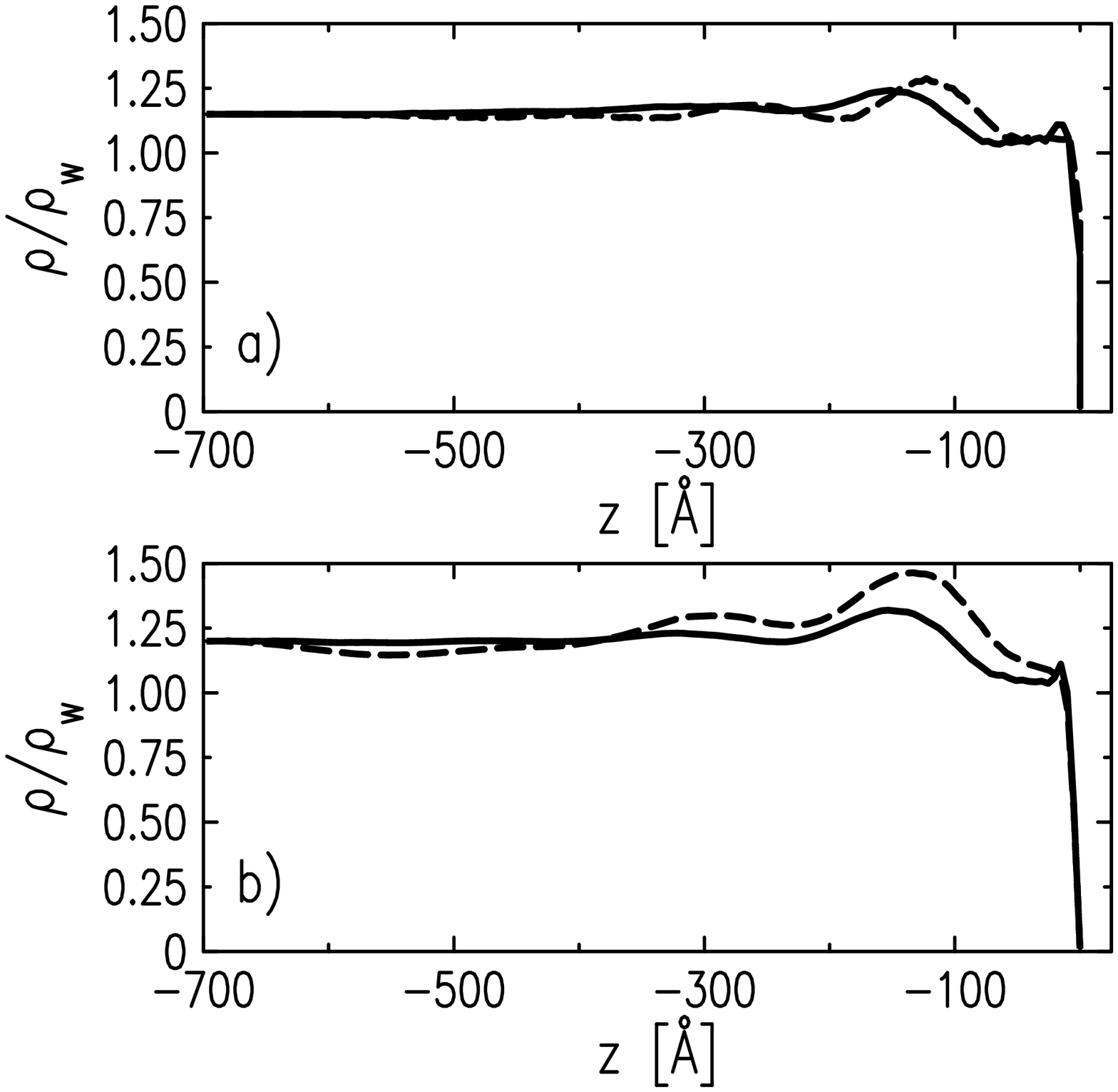, width=0.7\textwidth}

Figure 5. Reconstructed distribution profiles $\rho(z)$ divided by
the electron density in water $\rho(z)_w=0.333$\,{\it e$^-$}/\AA$^3$:
(à) the surface of the sol of 7-nm particles (dashed line)
immediately after the preparation of the sample and (solid
line) after $\sim 170$\,h; (b) the surface of the sol of 12-nm particles
(dashed line) immediately after the preparation of
the sample and (solid line) after $\sim 120$\,h.
\end{figure}

The model-independent approach is based on the
extrapolation of the asymptotic behavior of the $R(q_z)$ to the region
of high $q_z$ values under the assumption that the distribution
of the polarizability across the surface of the sample $\delta(z)$ or its derivatives
have jumps at the points $z_j$ [11, 12]. If all distances
between $z_j$ are different, only two physically
reasonable distributions $\delta(z)$ correspond to the reflection
coefficient $R(q_z)$ measured in a limited range of $q_z$.
These distributions differ from each other in the
arrangement of the points $z_j$ with respect to the
boundary of the substrate. Further, we chose the solution
at which the first point of discontinuity $z_1=0$
corresponds to the air–multilayer interface and the
other points $z_j$ are deep in the substance ($z_j<0$).
To describe the studied samples, it appeared to be sufficient
to consider only the points of discontinuity of the
zeroth and first orders, i.e., jumps in the function $\delta(z)$
and its derivative $\delta^\prime(z)$, respectively. The profiles $\delta(z)$
were numerically optimized by fitting the calculated
reflectivity curve (the function of $\delta(z)$ and $q_z$) to the
experimental data for $R(q_z)$ with the use of the standard
Levenberg–Marquardt algorithm \cite{28}.

For weakly absorbing substances, in the spectral
range of hard X rays, the model-independent depth
profiles of the electron density $\rho(z) \simeq \pi\delta(z)/(r_0\lambda^2)$,
where $r_0=2.814\times 10^{-5}$\,\AA{} is the classical radius of the
electron \cite{29}, can be calculated from the reconstructed
distributions of the optical constant $\delta(z)$.
Then, by comparing the reconstructed profile $\rho(z)$
with a certain structural model of the surface layer, the
specific area $A$ per structural unit  (ion, molecule, or
chemical group) in the layer with the thickness $d=z_2-z_1$
can be estimated as
\begin{equation}
A=\frac{\Gamma}{\int\limits_{z1}^{z2}\rho(z)dz},
\end{equation}
where $\Gamma$ is the number of electrons in the structural
unit. For example, for the DSPC molecule, $\Gamma =438$.

All dependencies $R(q_z)$ in Fig. 2 decrease as $\propto 1/q_z^4$. Further analysis shows that the function $\delta(z)$ has only
one singular point of the zeroth order. Figure 5à shows
the reconstructed profiles $\rho(z)$ in the surface of the layer of the sol of 7-nm particles divided by the
electron density in water under normal conditions $\rho(z)_w=0.333$\,{\it e$^-$}/\AA$^3$.
They qualitatively correspond to the model of electric double layer proposed in \cite{17,18}. The thickness of the densest layer (loose nanoparticle monolayer) with a maximum at a depth of $\sim 150$~\AA{} was $\approx 100$~\AA{} and approximately corresponds to
the diameter of SiO$_2$ nanoparticles in the sol. Immediately after the preparation of the sample (dashed line), the maximum electron density in this layer is
$\rho_{max} \approx 1.3\rho_w$, which is $\sim 12\%$ larger than the value in the bulk of the solution $\rho_b\approx 1.15\rho_w$; i.e., the concentration
of nanoparticles in this layer is larger than the
bulk value by a factor of $(\rho_{max}-\rho_w)/(\rho_b-\rho_w)\approx 2$.
The second layer with the same thickness and the concentration
of silicon dioxide particles exceeding the
bulk value by $\sim 20\%$ is simultaneously observed at a larger depth. The total thickness of the observed separation region reaches $\sim 500$~\AA{}.

After the aging of the sample for $\sim 170$\,h (solid line in Fig. 5a), the concentration of nanoparticles in the loose multilayer decreases by $\sim 20\%$, its position is
shifted deeper in the substrate, and the second layer
disappears. The total thickness of the separation layer
decreases to $\sim 300$~\AA{}, which is in agreement with the
estimate $\Lambda_D\approx 400$\,\AA{} for this solution. A narrow ($d_0\approx 20$~\AA{}) peak of the electron density is also
observed directly at the (air–silica sol) interface.

\begin{figure}
\hspace{0.5in}
\epsfig{file=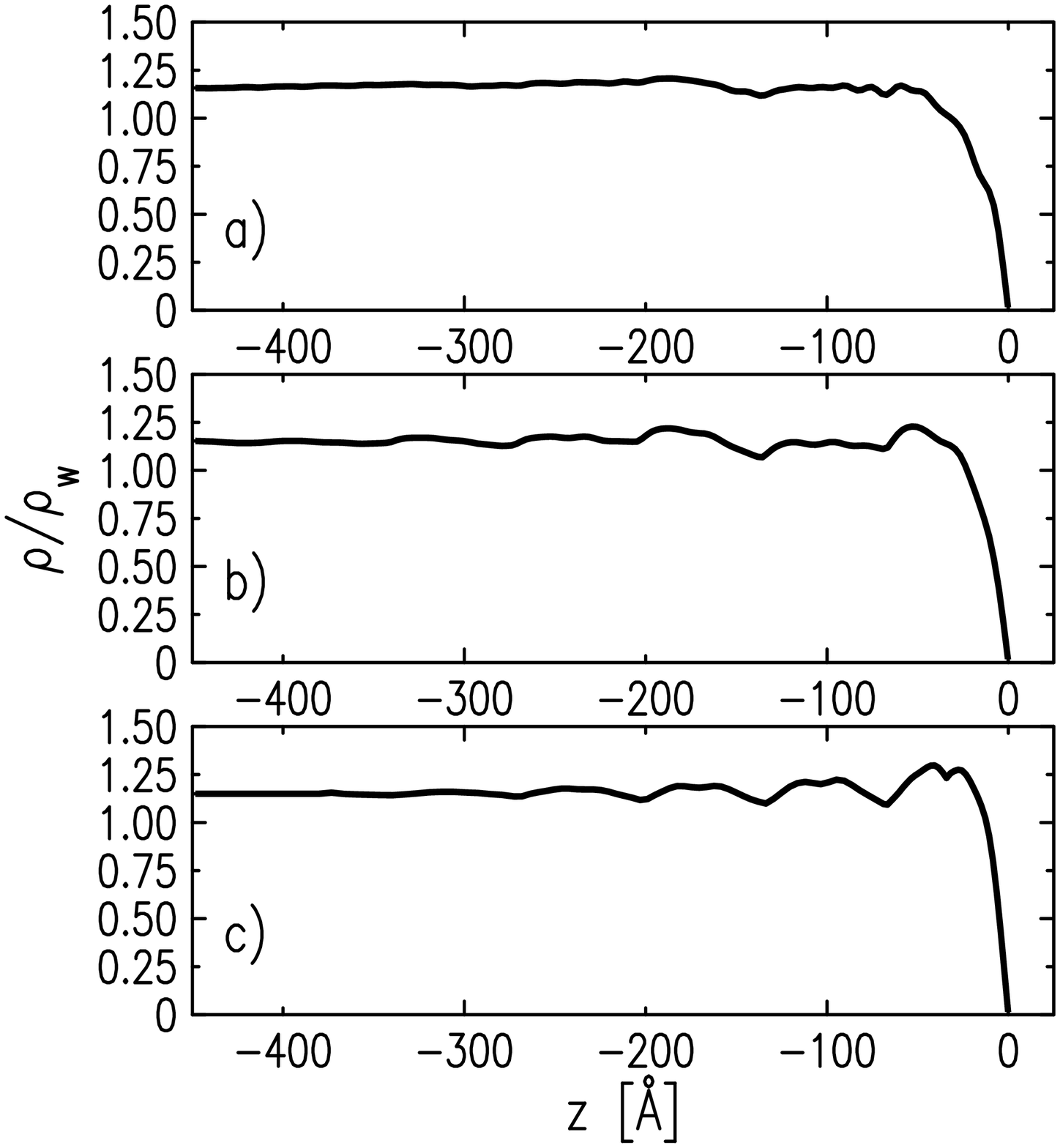, width=0.7\textwidth}

Figure 6. Reconstructed distribution profiles $\rho(z)$ divided by
the electron density in water $\rho(z)_w=0.333$\,{\it e$^-$}/\AA$^3$
for the DSPC lipid film on the surface of the sol of 7-nm
particles (a) 1, (b) 24, and (c) 96\,h after its deposition.
\end{figure}

The behavior of the electron density distributions
near the surface of the silica sol of 12-nm particles in
Fig. 5b is similar. Immediately after the preparation of
the sample (dashed line), the excess concentration of
SiO$_2$ nanoparticles in the loose monolayer of
nanoparticles is higher than the bulk value by a factor
of $\approx 2$. In addition, a depleted layer is present at a depth
of $\approx 550$~\AA{}, where $\rho(z)$ is $\sim 6\%$ lower than the value in
the bulk of this sol $\rho_b \approx 1.2\rho_w$. Thus, the concentration
of particles in this layer is $\sim 40\%$ lower than the
bulk value.

After $\sim 120$\,h (the solid line), the deep enriched and
depleted layers disappear and the concentration of
nanoparticles in the loose monolayer decreases to $\sim 1.4$ of the bulk value. The total thickness of the separation
region decreases from $\sim 700$\,\AA{} to $\sim 400$~\AA{}$\approx \Lambda_D$.
A thin layer with the thickness $d_0\sim 20$~\AA{} is also manifested on the surface.

Since all curves in Fig. 3 decrease as $\propto 1/q_z^6$, the further
consideration implies the presence of the points
of discontinuity of the first order in the structure. The
reconstructed depth distributions of the electron density $\rho(z)$
divided by $\rho_w$ are shown in Fig. 6.

\begin{figure}
\hspace{0.5in}
\epsfig{file=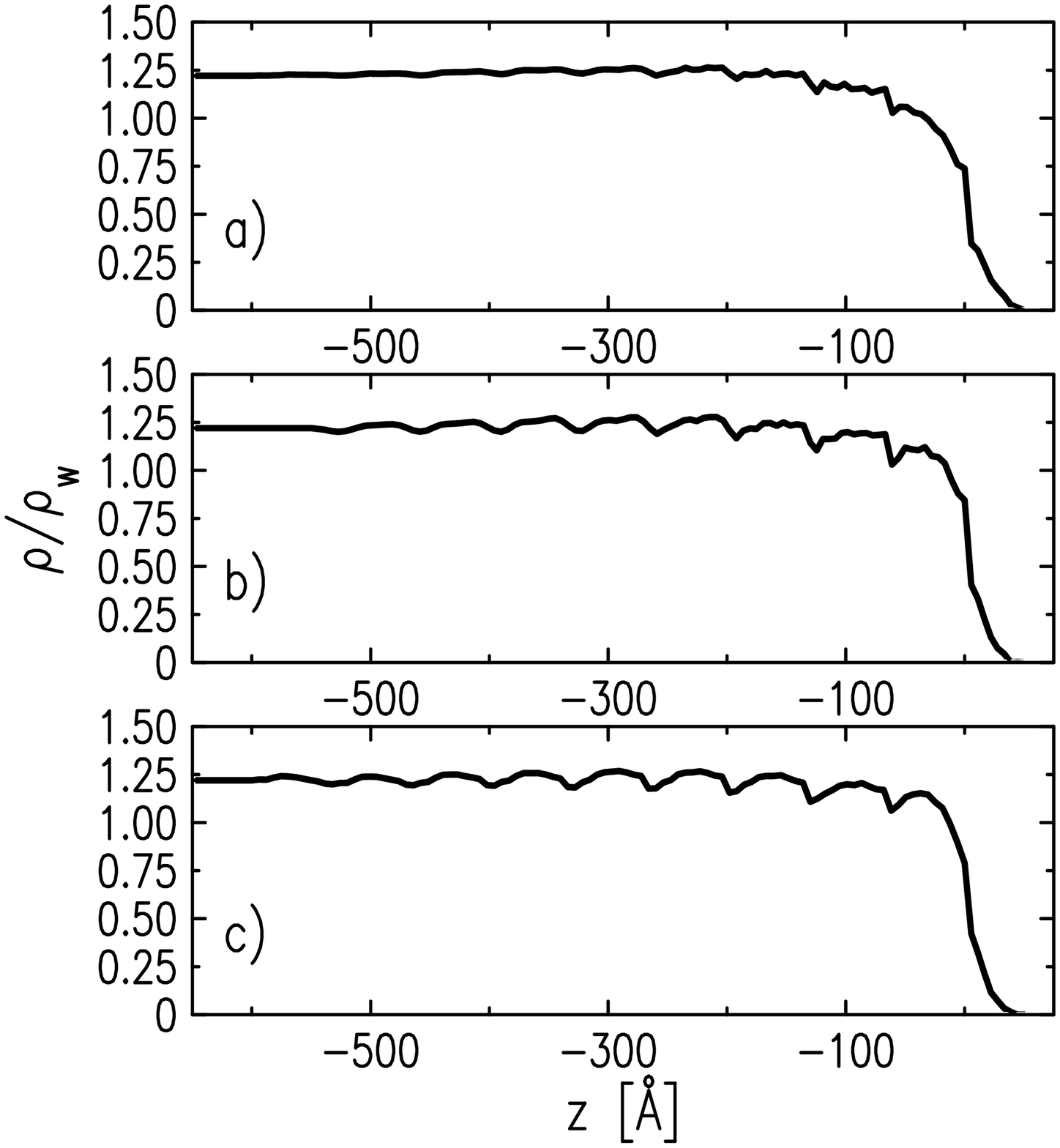, width=0.7\textwidth}

Figure 7. Reconstructed distribution profiles $\rho(z)$ divided by
the electron density in water $\rho(z)_w=0.333$\,{\it e$^-$}/\AA$^3$
for the DSPC lipid film on the surface of the sol of 12-nm
particles (a) 24, (b) 70, and (c) 200\,h after its deposition.
\end{figure}

In 1 h after preparation (Fig. 6a), the electron density
at a depth up to 350\,\AA{} ($\sim \Lambda_D$) is higher than the bulk value by 3-4\% and the lipid film on the surface of the silica sol is apparently in a disordered state. The comparison of the structure of the surface layer of this sample with the above-considered surface of the sol indicates the disappearance of the feature associated with the loose monolayer of nanoparticles; i.e., the spreading of the lipid film is accompanied by a significant redistribution of particles.
After aging of the sample for 24 h, the reconstructed
electron density distribution (Fig. 6b) at
depth up to 350\,\AA{} exhibits a quasiperiodic structure
with a characteristic dimension of $d\approx 68$\,\AA{}, which
corresponds to the double length of the DSPC molecule
($\approx 35$\,\AA).

Finally, after $\sim 100$\,h, the profile $\rho(z)$ (Fig. 6c)
demonstrates four pronounced lipid bilayers with the
thickness $d = 68.1\pm 0.9$\,\AA{}. The depth distribution
of the electron density inside each bilayer is symmetric,
which indicates a good ordering of molecules in it.

The angular dependences in Fig. 4 decrease as $\propto 1/q_z^4$; for this reason, the structure was reconstructed
under the assumption of the zeroth order of points of
discontinuity in the distribution of the polarizability.
All reconstructed electron density distributions in
Fig. 7 exhibit a structure of six to eight layers with a
characteristic period of ($66.7\pm 2.0$)\,\AA{}, which corresponds
to the thickness of the DSPC bilayer. The total
thickness of the multilayer exceeds 500\,\AA{}. A disordered layer with a low density and a thickness of about
the DSPC monolayer $\approx 40$\,\AA{} is also present on the surface.

One day after the preparation of the sample
(Fig. 7a), the electron density in bilayers near the air– phospholipid interface is significantly lower than that
in bilayers near the silica sol substrate and its distribution inside individual bilayers is strongly asymmetric,
which indicates their incomplete ordering. The area
per lipid molecule for the most ordered bilayers at a
depth of 250--400\,\AA{} is estimated as $A=(37\pm 2)$\,\AA{}$^2$.

After $\sim 70$\,h (Fig. 7b), the integral electron density of bilayers hardly varies ($A = (36 \pm 2)$\,\AA{}$^2$), but the
interfaces between them become more pronounced.
This indicates the lateral ordering of the structure of
the lipid film.

Finally, after $\sim 200$\,h (Fig. 7c), the electron density
distribution within individual bilayers becomes symmetric
at an unchanged integral electron density of the
entire structure.

Thus, according to the reported data, a macroscopically
flat structure is formed on the pure (air–silica
sol) interface. The electron density profiles reconstructed with the model-independent approach are in
agreement with the concept of the structure of the
electric double layer on the surface of the hydrosol \cite{17,18}. The gradient of the surface potential on it appears owing to the difference between the potentials of "electric image" forces for Na$^+$ cations and silica
nanoparticles with a large negative charge ($\sim 10^3$ electrons)
(macroions). Within several days of aging of the
sample, nanoparticles in the surface layer of the silica
sol are redistributed and a loose monolayer is formed.
As a result, the distance between the plane of the closest
approach of anion particles and the surface is
$\sim 100$\,\AA, whereas Na$^+$ cations are accumulated immediately on the interface in a thin layer of the space charge with the thickness $d_0\sim 20$\,\AA \cite{30,31}.

Our data also demonstrate the ordering with the
time in the DSPC phospholipid multilayer deposited
on the silica sol substrate. The characteristic thickness
of the formed structure is $\sim \Lambda_D$, as in the case of the
clean surface of the hydrosol.

During the first hour after the deposition of lipid, a
thin layer with the thickness $\approx 130$\,\AA{}, which is separated
from the substrate by a disordered material film
with the thickness $\sim 200$\,\AA{}, is formed at the hydrosol–air interface. In this case, the feature on the reflectivity
curve that is due to the loose monolayer of nanoparticles
almost completely disappears. This indicates a
significant rearrangement of the surface. Such a structure
can be explained under the assumption that, e.g.,
the thin layer consists of three "liquid" lipid bilayers
with the area per molecule $A = (55.7 \pm 0.7)$\,\AA{}$^2$. This $A$
value is in agreement with, e.g., an estimate for the
area per molecule in the bilayer walls of vesicles \cite{4}.

During the next day, a quasiperiodic multilayer
structure including five to eight partially ordered lipid
bilayers is formed. Finally, several days ($(1-7)\times 10^{5}$\,s)
after the preparation of the sample, the structure of the
reconstructed profiles of the lipid multilayer does not
change and a fine structure of monolayers can be
identified inside individual bilayers. The average area
per molecule in multilayers is $A = (36 \pm 2)$\,\AA{}$^2$, which
is noticeably smaller than its estimate $A_{0}=(41.6 \pm 0.7)$\,\AA{}$^2$ from diffraction data for crystalline
bilayers \cite{2,6}. The comparison of $A$ with $A_0$ gives the
excess electron density averaged over the multilayer
$\Gamma(A_0-A)/A_0 = 60\pm 20$ electrons per lipid molecule,
which corresponds to four to eight Na$^+$ ions and H$_2$O
molecules per DSPC molecule. The depth dependence
of the distribution $\rho(z)$ for bilayers can be due
both to filling defects and to inhomogeneous accumulation of Na$^+$ and H$_2$O in layers.

It is noteworthy that a monolayer is formed on the
surface of the silica sol in the case of a low surface concentration
of DSPC. In this case, the distance from
the plane of closest approach between nanoparticles to
the surface decreases to the thickness of the monolayer
$\approx 35$\,\AA{} and nanoparticles are condensed on its hydrophilic
surface, at which the surface concentration of particles is higher than the bulk value by a factor of $\sim 2$
\cite{18}. In the considered case of a high surface concentration
of DSPC molecules, the position of the plane
of the closest approach of silica particles with respect
to the surface is given by the thickness of the multilayer $\sim \Lambda_d$.
In this case, any pronounced interface between
it and the loose monolayer of nanoparticles is not
observed, which indicates a decrease in their surface
concentration.

A number of previous molecular dynamics calculations
show that Na$^+$ ions can be introduced in phospholipid
membranes, thus forming a positive surface
potential [32-34]. However, such a mechanism can
explain the excess electron density only in the bilayer
directly adjacent to the substrate rather than in the
entire multilayer.

An analogy can be seen between the formation of
the surface structure with a transient process in an $RC$
circuit, where the capacitance of the electric double
layer $C$ and the resistance of the multilayer $R$ are connected in parallel to the source of the current of Na$^+$
ions, which is generated by electric image forces. The
characteristic time of charging the capacitance is
$\tau = RC \sim 10^5 - 7 \cdot 10^5$\,s, where $C \sim \epsilon_0\epsilon_1/\Lambda_D$ and $R \sim \rho\Lambda_D$
are the Helmholtz capacitance and the
resistance of the multilayer per unit area, respectively.
Consequently, the resistance of the DSPC bilayer per
unit area is $\rho d   \sim \tau d/(\epsilon_0\epsilon_1) \approx 4\cdot 10^{7} - 3 \cdot 10^{8}$\,Ohm$\cdot$m$^2$ at the
static dielectric constant of the multilayer $\epsilon_1 \approx 2$. This value is much lower than the resistance
$10^{10} - 10^{13}$\,Ohm$\cdot$m$^2$ previously obtained from the measurements of the ionic conductivity of unmodified phospholipid membranes \cite{35}. In other words, the time $\tau$ for the observed structure is two to five orders of magnitude smaller than the expected value.

The electric field near the surface of the loose
monolayer of nanoparticles, which orients the dipoles
of DSPC molecules at the initial time, reaches $E > 10^{9}$\,V/m \cite{31}. The voltage drop $\Delta V$ across the
thickness of the bilayer is $\Delta V= Ed > 7$\,V; i.e., the
condition for its electric instability or electroporation
is certainly satisfied ($\Delta V \geq 0.1$\,V) [36-39]. In this case, a certain porous structure is apparently formed
in the multilayer through which the transport of Na
ions from the bulk of the hydrosol to the interface with
air occurs more efficiently as compared to ohmic conductivity
[40, 41]. In our opinion, such a mechanism
of charge transport explains both a relatively small
value and a high electron density in bilayers.

We are grateful to I.\,V.\,Kozhevnikov for stimulating
discussions of the experimental results. This work was
supported in part by the Russian Foundation for Basic
Research (project no. 15-32-20935).

\end{document}